\documentclass{jpsj2}
\title{Magnetic Phase Diagram of Spin-1/2 Two-Leg Ladder 
with Four-Spin Ring Exchange}

\author{Toshiya {\sc Hikihara} and Shoji {\sc Yamamoto}}

\inst{Department of Physics, Hokkaido University, Sapporo 060-0810}

\abst{
We study the spin-$1/2$ two-leg Heisenberg ladder 
with four-spin ring exchanges under a magnetic field.
We introduce an exact duality transformation which is an extension of 
the spin-chirality duality developed previously 
and yields a new self-dual surface in the parameter space.
We then determine the magnetic phase diagram using the numerical approaches 
of the density-matrix renormalization-group and exact diagonalization methods.
We demonstrate the appearance of a magnetization plateau 
and the Tomonaga-Luttinger liquid 
with dominant vector-chirality quasi-long-range order 
for a wide parameter regime of strong ring exchange.
A ^^ ^^ nematic" phase, in which magnons form bound pairs 
and the magnon-pairing correlation functions dominate, 
is also identified.
}

\kword{
ring exchange, spin ladder, frustration, vector chirality, nematic phase,
density-matrix renormalization-group method
}

\begin{document}
\maketitle

\section{Introduction}\label{sec:intro}
Multiple-spin ring-exchange process -- a process in which 
several particles with spin permute their positions in a cyclic fashion --
has been found to be important in a wide variety of materials.
It has been established that ring exchanges are responsible 
for the magnetism of solid $^3$He\cite{RogerHD1983,Roger1984}
as well as of two-dimensional quantum solids including 
$^3$He adsorbed on graphite\cite{GreywallB1989,Roger1990,RogerBBCG1998} 
and Wigner crystals.\cite{Roger1984,OkamotoK1998,KatanoH2000,BernuCC2001}
Ring exchanges were also found to be relevant to 
some strongly correlated electron systems 
such as the two-leg spin ladder compound 
La$_x$Ca$_{14-x}$Cu$_{24}$O$_{41}$\cite{BrehmerMMNU1999,MatsudaKEBM2000
,MatsudaKEBM2000a,SchmidtKU2001,NunnerBKWG2002}
and two-dimensional antiferromagnet La$_2$CuO$_4$.\cite{HondaKW1993
,MizunoTM1999,MizunoTM2001,Coldea2001,KataninK2002,KataninK2003}
It has been suggested recently that ring exchanges can be dominant 
in quantum wires with a shallow confining potential.\cite{KMM2006,KMHM2007}

In contrast to two- and three-spin exchange processes, 
the ring exchanges involving more than three spins take 
a very complicated form in terms of spin operators 
including bilinear, biquadratic, and higher-order couplings in general, 
which compete with each other.
The multiple-spin ring exchanges therefore contain strong frustration 
in themselves and can induce exotic phenomena.
Thus motivated, extensive studies have been devoted to clarifying
the effects of ring exchanges on quantum spin systems 
in two-leg ladder,\cite{HondaH2001,HijiiN2002,MullerVM2002
,LaeuchliST2003,HikiharaMH2003,MomoiHNH2003,GritsevNB2004}
square,\cite{ChubukovGB1992,LaeuchliDLST2005,ShannonMS2006}
and triangular lattices.\cite{MomoiKN1997,KuboM1997
,MisguichBLW1998,MisguichLBW1998,MomoiSS2006}
It has been revealed that the four-spin ring exchange 
can lead to novel quantum states, e.g., 
the scalar-chirality ordered state in the two-leg ladder 
system,\cite{LaeuchliST2003,HikiharaMH2003,MomoiHNH2003}
the vector-chirality\cite{LaeuchliDLST2005} and 
nematic\cite{ShannonMS2006} orderings in the square lattice, 
and the octapolar order\cite{MomoiSS2006} in the triangular lattice.

In this paper, we study the $S=1/2$ two-leg Heisenberg ladder system with four-spin ring exchanges under a magnetic field.
The magnetic field lowers the symmetry of the model from SU(2) to U(1), 
and as a result, opens a possibility of novel quantum states.
The model was studied previously by numerical exact-diagonalization method 
and perturbation theory and it was predicted that the model exhibited 
a plateau in the magnetization curve\cite{SakaiH1999,NakasuTHOS2001}
and a field-induced incommensurate quasi long-range order (LRO)
around the plateau.\cite{SakaiO2005}
Further, it was also suggested very recently that 
the interplay between four-spin exchanges and the magnetic field 
might give rise to a true LRO of vector chirality.\cite{Sato2007}
Nevertheless, these studies were limited to antiferromagnetic 
bilinear couplings and rather weak ring exchange, 
and therefore, the properties of the model for 
the regime of strong ring exchange and/or of ferromagnetic 
bilinear coupling remain unclear.

The aim of the paper is to investigate the ground-state properties 
of the ladder system with four-spin ring exchanges 
for the whole parameter space of both the antiferromagnetic and ferromagnetic 
bilinear exchanges and a finite magnetic field.
We first employ a theoretical approach 
using duality transformations.
The so-called spin-chirality duality 
transformation\cite{HikiharaMH2003,MomoiHNH2003} 
has proven to be a powerful tool to study 
the ladder system with four-spin exchanges.
We further introduce another duality transformation, 
which is a simple extension of the spin-chirality duality 
but gives a nontrivial duality mapping on coupling constants 
in the Hamiltonian as well as on various order parameters.
In particular, applying the latter transformation to the ladder 
with the ring exchange yields a new self-dual line, 
which divides the parameter region of strong ring exchange 
from that of strong ferromagnetic bilinear exchange.

Moreover, we study the system using numerical approaches 
of the density-matrix renormalization-group (DMRG) 
and exact diagonalization methods 
and determine the magnetic phase diagram.
We then directly observe the magnetization plateau 
appearing for a wide parameter region of strong ring-exchange coupling.
We find various critical states 
with different dominant quasi-LRO: 
In addition to the critical state 
with dominant N\'{e}el-type-spin quasi-LRO
found in usual antiferromagnetic two-leg ladder without ring exchange,
the vector-chirality-dominant critical state appears 
for strong ring exchange.
More interestingly, we show that 
a nematic phase, in which magnons form bound pairs and 
the magnon-pairing correlations are dominant, 
emerges in a finite region adjacent to the ferromagnetic phase.

The paper is organized as follows.
In Section \ref{sec:model}, we introduce the model Hamiltonian 
and summarize its known properties obtained from previous studies.
The duality transformations and their application to 
the ladder system with four-spin exchanges are discussed 
in Section \ref{sec:duality}.
Section \ref{sec:numerics} shows 
our numerical results on the magnetic phase diagram. 
The results on the magnetization plateau and 
the vector-chirality-dominant critical phase
appearing in the region of strong ring exchange 
are respectively presented 
in Section \ref{subsec:plateau} and \ref{subsec:vectorchirality}.
The nematic phase in the regime of 
strong ferromagnetic bilinear coupling is discussed 
in Section \ref{subsec:nematic}.
Section \ref{sec:summary} contains summary and concluding remarks.

\section{Model}\label{sec:model}
We consider the $S=1/2$ Heisenberg spin ladder with four-spin ring-exchanges 
under a magnetic field.
The model Hamiltonian is given by
\begin{eqnarray}
\mathcal{H} &=& \mathcal{H}_0 + \mathcal{H}_h
\label{eq:Ham} \\
\mathcal{H}_0 &=& J_{\rm rung} \sum_l {\bf s}_{1,l} \cdot {\bf s}_{2,l}
%\nonumber \\
%&&
+ J_{\rm leg} \sum_l \left( {\bf s}_{1,l}\cdot{\bf s}_{1,l+1} 
                             + {\bf s}_{2,l}\cdot{\bf s}_{2,l+1} \right)
\nonumber \\
&&+ J_4 \sum_l ( P_l + P_l^{-1} )
\label{eq:Ham_0} \\
\mathcal{H}_h &=& - h \sum_l \left( s^z_{1,l} + s^z_{2,l} \right),
\label{eq:Ham_Zeeman}
\end{eqnarray}
where ${\bf s}_{j,l}$ is a spin-$1/2$ operator 
at a site on $j$-th leg and $l$-th rung 
and $J_{\rm rung}$ ($J_{\rm leg}$) denotes the bilinear exchange
constant on rungs (legs).
The operator $P_l$ ($P_l^{-1}$) represents the four-spin ring-exchange process
in which four spins on a plaquette $\{(1,l),(2,l),(2,l+1),(1,l+1)\}$ 
exchange their positions clockwise (counterclockwise).
Throughout this paper, we consider the case of 
$J_{\rm rung} = J_{\rm leg} = J$ and $J_4 \ge 0$ and 
parameterize them as 
\begin{equation}
J = \cos\theta,~~~~ J_4 = \sin\theta, 
\label{eq:couplingparameter}
\end{equation}
where $0 \le \theta \le \pi$.

In terms of the spin operators, the ring exchange is expressed 
in a complicated form,
\begin{eqnarray}
P_l + P_l^{-1}
&=& {\bf s}_{1,l}\cdot{\bf s}_{2,l} + {\bf s}_{1,l+1}\cdot{\bf s}_{2,l+1}
%\nonumber\\
%&&
+ {\bf s}_{1,l}\cdot{\bf s}_{1,l+1} + {\bf s}_{2,l}\cdot{\bf s}_{2,l+1} 
%\nonumber \\
%&&
+ {\bf s}_{1,l}\cdot{\bf s}_{2,l+1} + {\bf s}_{2,l}\cdot{\bf s}_{1,l+1}
\nonumber\\
&&+ 4({\bf s}_{1,l} \cdot {\bf s}_{2,l})
     ({\bf s}_{1,l+1} \cdot {\bf s}_{2,l+1}) 
%\nonumber \\
%&&
+ 4({\bf s}_{1,l} \cdot {\bf s}_{1,l+1})
     ({\bf s}_{2,l} \cdot {\bf s}_{2,l+1})  
\nonumber \\
&&- 4({\bf s}_{1,l} \cdot {\bf s}_{2,l+1}) 
     ({\bf s}_{2,l} \cdot {\bf s}_{1,l+1}), 
\label{eq:P4}
\end{eqnarray}
where we omit a constant term.
Using this expression, the Hamiltonian (\ref{eq:Ham}) is rewritten 
in a form containing bilinear and biquadratic couplings.
For the later use, we introduce an extended Hamiltonian 
with generalized four-spin exchanges defined as
\begin{eqnarray}
\mathcal{H}_{\rm ext} &=&
J_{\rm r} \sum_l {\bf s}_{1,l} \cdot {\bf s}_{2,l}
%\nonumber \\
%&&
+ J_{\rm l} \sum_l \left( {\bf s}_{1,l} \cdot {\bf s}_{1,l+1} 
                          + {\bf s}_{2,l} \cdot {\bf s}_{2,l+1} \right)
\nonumber \\
&&+ J_{\rm d} \sum_l \left( {\bf s}_{1,l} \cdot {\bf s}_{2,l+1}
                          + {\bf s}_{2,l} \cdot {\bf s}_{1,l+1} \right)
\nonumber \\
&&+ J_{\rm rr} \sum_l \left({\bf s}_{1,l  } \cdot {\bf s}_{2,l  } \right)
                      \left({\bf s}_{1,l+1} \cdot {\bf s}_{2,l+1} \right)
%\nonumber\\
%&&
+ J_{\rm ll} \sum_l \left({\bf s}_{1,l  } \cdot {\bf s}_{1,l+1} \right)
                      \left({\bf s}_{2,l  } \cdot {\bf s}_{2,l+1} \right)
\nonumber \\
&&+ J_{\rm dd} \sum_l \left({\bf s}_{1,l  } \cdot {\bf s}_{2,l+1} \right)
                      \left({\bf s}_{2,l  } \cdot {\bf s}_{1,l+1} \right)
\nonumber \\
&&- h \sum_l \left( s^z_{1,l} + s^z_{2,l} \right).
\label{eq:Ham_ext}
\end{eqnarray}
The ring-exchange model (\ref{eq:Ham}) 
with Eq.\ (\ref{eq:couplingparameter}) 
is expressed as a parameter line 
in the extended parameter space, 
$J_{\rm r} = J + 2 J_4$, $J_{\rm l} = J + J_4$, $J_{\rm d} = J_4$, and 
$J_{\rm rr} = J_{\rm ll} = -J_{\rm dd} = 4 J_4$.

The ground-state properties of the ring-exchange model (\ref{eq:Ham}) 
have been studied for some limiting cases.
For zero magnetic field $h = 0$, 
the ground-state phase diagram for the whole parameter region 
was determined by extensive numerical calculations.\cite{LaeuchliST2003}
It was shown that besides 
the conventional rung-singlet phase ($\theta < 0.07\pi$) 
and ferromagnetic phase ($0.94\pi < \theta$) 
the model exhibited unconventional phases, including 
the staggered-dimer phase ($0.07 \pi < \theta < 0.15\pi$), 
the scalar-chirality phase ($0.15\pi < \theta < 0.39\pi$), 
and the singlet phases 
with dominant vector-chirality ($0.39\pi < \theta < 0.85\pi$) 
and collinear-spin ($0.85\pi < \theta < 0.94\pi$) short-range orders.
All the phases except for the ferromagnetic one 
have a finite energy gap.
Furthermore, the spin-chirality duality 
transformation,\cite{HikiharaMH2003,MomoiHNH2003} 
which applies also to the case of $h > 0$, 
revealed various duality relations between order parameters as well as 
duality mappings in the model Hamiltonian.
We will discuss the duality transformation in the following section.

For the antiferromagnetic two-leg spin ladders without ring exchanges, 
Eq.\ (\ref{eq:Ham}) with $J > 0$ and $J_4 = 0$, 
the effect of the magnetic field has been well understood.\cite{ChitraG1997
,UsamiS1998,GiamarchiT1999,FurusakiZ1999,HikiharaF2001}
The energy gap above the singlet ground state at $h = 0$ decreases 
as the field $h$ increases and closes at a critical field $h_{\rm c}$.
The system then enters a critical regime, 
where the low-energy excitations are described 
by the one-component Tomonaga-Luttinger (TL) 
liquid.\cite{ChitraG1997,GiamarchiT1999,FurusakiZ1999}
Throughout the regime, the transverse spin correlation with 
$(q_x,q_y)=(\pi, \pi)$ dominates.
The field-dependence of the TL-liquid parameter, 
which governs the low-energy physics of the system, 
as well as that of nonuniversal coefficients included in the TL-liquid theory 
were also obtained numerically.\cite{UsamiS1998,HikiharaF2001}
When $h$ exceeds the saturation field $h_{\rm s}$, 
the fully polarized ground state emerges.

Compared with the cases above, little is known 
about the case with $J_4 > 0$ and $h > 0$.
Previous studies were performed for 
antiferromagnetic bilinear coupling $J > 0$ 
and rather small ring exchange $J_4 \lesssim J$.
From numerical exact-diagonalization studies for small clusters 
with the level-spectroscopy analysis, 
it has been shown that for not too small $J_4$ 
a plateau emerges in the magnetization curve 
at a half of the saturated magnetization.\cite{SakaiH1999,NakasuTHOS2001}
The perturbation theory around the strong-rung-coupling limit 
$J_{\rm rung} \to \infty$ have concluded that 
the ground state at the plateau breaks the translational symmetry, 
resulting in the LRO of the staggered pattern of 
spin-singlet and spin-triplet rungs.
A phenomenon called ^^ ^^ $\eta$-inversion", 
in which a longitudinal incommensurate spin correlation 
becomes stronger than the transverse-spin one,
was predicted by exact diagonalization, 
while any direct observation of the behavior of the correlation functions 
has not been achieved yet.\cite{SakaiO2005} 
The possibility of the true LRO of vector-chirality 
was also discussed within the bosonization approach\cite{Sato2007}
though it is still elusive whether or not it realizes.

\section{Duality relations}\label{sec:duality}
In this section, we discuss two duality transformations 
in the extended Hamiltonian $\mathcal{H}_{\rm ext}$ in Eq.\ (\ref{eq:Ham_ext}) 
and the corresponding duality relations between various phases.
The one transformation is the so-called spin-chirality duality
and the other is its extension.
The former was already introduced and discussed 
in Refs.\ \citen{HikiharaMH2003} and \citen{MomoiHNH2003} in detail,
and we present here a brief summary for completeness.

The spin-chirality duality transformation is defined as 
a mapping from original spin operators in a rung, 
${\bf s}_{1,l}$ and ${\bf s}_{2,l}$, into new pseudospin ones; 
it is given by
\begin{eqnarray}
\tilde{\bf S}_{1,l} &\equiv&
 \frac{1}{2} \left( {\bf s}_{1,l} + {\bf s}_{2,l} \right)
                   - {\bf s}_{1,l} \times {\bf s}_{2,l},
\nonumber \\
\tilde{\bf S}_{2,l} &\equiv&
 \frac{1}{2} \left( {\bf s}_{1,l} + {\bf s}_{2,l} \right)
                   + {\bf s}_{1,l} \times {\bf s}_{2,l}.
\label{eq:duality1}
\end{eqnarray}
The new pseudospin operators obey the usual commutation relations for spins 
and satisfy $(\tilde{\bf S}_{1,l})^2 = (\tilde{\bf S}_{2,l})^2 = 3/4$.
The spin-chirality transformation gives 
duality mappings between various order parameters.
For example, the $(q_x, \pi)$-spin order operator,
\begin{equation}
\mathcal{O}_{\rm s} (q_x, \pi) 
= \sum_l e^{iq_x l} \left({\bf s}_{1,l} - {\bf s}_{2,l} \right),
\label{eq:Ospipi}
\end{equation}
converts into the vector-chirality one,
\begin{equation}
\tilde{\mathcal{O}}_{\rm vc}(q_x)
= 2 \sum_l e^{iq_x l} \tilde{\bf S}_{1,l} \times \tilde{\bf S}_{2,l},
\label{eq:Ovc}
\end{equation}
and vise versa (up to an overall sign factor),
while the total rung-spin operator ${\bf s}_{1,l} + {\bf s}_{2,l}$ 
is self-dual under the transformation.
We note that the duality transformation (\ref{eq:duality1}) 
corresponds to a unitary transformation,\cite{MomoiHNH2003}
\begin{equation}
\tilde{\bf S}_{n,l} 
= U(\{ \theta_l \}) {\bf s}_{n,l} U^\dagger(\{ \theta_l \}),
\label{eq:unitary1}
\end{equation}
with a unitary operator, 
\begin{equation}
U(\{ \theta_l \}) = \prod_l \exp\left[ i \theta_l 
\left({\bf s}_{1,l}\cdot{\bf s}_{2,l} - \frac{1}{4} \right) \right],
\label{eq:unit_spin}
\end{equation}
and $\theta_l = \pi/2$ for arbitrary $l$.
In this paper, we call this transformation the duality {\it I}.

When applied to the extended model $\mathcal{H}_{\rm ext}$,
the duality transformation {\it I} 
leaves the form of the Hamiltonian unchanged 
and gives a duality mapping of the coupling parameters.
To see the parameter mapping, it is convenient to 
rewrite the Hamiltonian as
\begin{eqnarray}
\mathcal{H}_{\rm ext} &=& 
J_{\rm r} \sum_l {\bf s}_{1,l} \cdot {\bf s}_{2,l}
%\nonumber \\
%&&
+ J_{\rm rr} \sum_l \left({\bf s}_{1,l  } \cdot {\bf s}_{2,l  } \right)
                      \left({\bf s}_{1,l+1} \cdot {\bf s}_{2,l+1} \right)
\nonumber \\
&&+ W \sum_l \left({\bf s}_{1,l  } + {\bf s}_{2,l } \right) \cdot
             \left({\bf s}_{1,l+1} + {\bf s}_{2,l+1} \right) 
\nonumber \\
&&+ X \sum_l \left[ 
             \left({\bf s}_{1,l  } \cdot {\bf s}_{1,l+1} \right)
             \left({\bf s}_{2,l  } \cdot {\bf s}_{2,l+1} \right)
           + \left({\bf s}_{1,l  } \cdot {\bf s}_{2,l+1} \right)
             \left({\bf s}_{1,l+1} \cdot {\bf s}_{2,l  } \right) \right]
\nonumber\\
&&+ Y \sum_l \left[ 
             \left({\bf s}_{1,l  } - {\bf s}_{2,l  } \right) \cdot
             \left({\bf s}_{1,l+1} - {\bf s}_{2,l+1} \right)
         + 4 \left({\bf s}_{1,l  } \times {\bf s}_{2,l  } \right) \cdot
             \left({\bf s}_{1,l+1} \times {\bf s}_{2,l+1} \right) \right]
\nonumber \\
&&+ Z \sum_l \left[ 
             \left({\bf s}_{1,l  } - {\bf s}_{2,l  } \right) \cdot
             \left({\bf s}_{1,l+1} - {\bf s}_{2,l+1} \right)
         - 4 \left({\bf s}_{1,l  } \times {\bf s}_{2,l  } \right) \cdot
             \left({\bf s}_{1,l+1} \times {\bf s}_{2,l+1} \right) \right]
\nonumber \\
&&- h \sum_l \left(s^z_{1,l  } + s^z_{2,l } \right),
\label{eq:H_ext2}
\end{eqnarray}
where
\begin{eqnarray}
W&=&\frac{1}{2}(J_{\rm l}+J_{\rm d}),~~~~~
X=\frac{1}{2}(J_{\rm ll}+J_{\rm dd}),
\nonumber \\
Y&=&\frac{1}{16}(J_{\rm ll}-J_{\rm dd})+\frac{1}{4}(J_{\rm l}-J_{\rm d}),
\nonumber \\
Z&=&-\frac{1}{16}(J_{\rm ll}-J_{\rm dd})+\frac{1}{4}(J_{\rm l}-J_{\rm d}).
\end{eqnarray}
After some algebra, one finds that the duality transformation 
changes the sign of the coupling constant $Z$ to $-Z$ 
and leaves the other terms unchanged.
Therefore, the duality {\it I} 
leads to the mapping in the seven-dimensional parameter space, 
$( J_{\rm r}, J_{\rm rr}, W, X, Y, Z, h)$ 
to $( J_{\rm r}, J_{\rm rr}, W, X, Y, -Z, h)$.
In terms of the original coupling parameters, the mapping is given by
\begin{eqnarray}
\tilde{J}_{\rm r}&=& J_{\rm r},~~~~~\tilde{J}_{\rm rr}=J_{\rm rr},
\nonumber \\
\tilde{J}_{\rm l}&=& \frac{1}{2}(J_{\rm l}+J_{\rm d})
+\frac{1}{8}(J_{\rm ll}-J_{\rm dd}),
%\nonumber \\
~~~~~
\tilde{J}_{\rm d} =  \frac{1}{2}(J_{\rm l}+J_{\rm d})
-\frac{1}{8}(J_{\rm ll}-J_{\rm dd}),
\nonumber \\
\tilde{J}_{\rm ll}&=& 2(J_{\rm l}-J_{\rm d})
+\frac{1}{2}(J_{\rm ll}+J_{\rm dd}),
%\nonumber \\
~~~~~
\tilde{J}_{\rm dd} = -2(J_{\rm l}-J_{\rm d}) 
+ \frac{1}{2}(J_{\rm ll}+J_{\rm dd}).
\end{eqnarray}
The extended model $\mathcal{H}_{\rm ext}$ with $Z = 0$, 
equivalently $4(J_{\rm l} -J_{\rm d}) = J_{\rm ll} -J_{\rm dd}$, 
is therefore self-dual under the duality transformation {\it I}.

Since the duality transformation {\it I} 
[Eq.\ (\ref{eq:duality1}) or Eq.\ (\ref{eq:unitary1})] 
consists of independent transformations within each rung, 
one can construct arbitrary transformations 
with any set of the phase factors $\{ \theta_l \}$.
Among them, a set $\{ \theta_l = (-1)^l \pi/2 \}$ 
yields another useful duality transformation given by
\begin{eqnarray}
\bar{\bf S}_{1,l} &\equiv&
 \frac{1}{2} \left( {\bf s}_{1,l} + {\bf s}_{2,l} \right)
                   - (-1)^l {\bf s}_{1,l} \times {\bf s}_{2,l},
\nonumber \\
\bar{\bf S}_{2,l} &\equiv&
 \frac{1}{2} \left( {\bf s}_{1,l} + {\bf s}_{2,l} \right)
                   + (-1)^l {\bf s}_{1,l} \times {\bf s}_{2,l}.
\label{eq:duality2}
\end{eqnarray}
We note that the transformation is 
a product of the duality {\it I} 
and the exchange of two spins in every second rung, 
${\bf s}_{1,l} \leftrightarrow {\bf s}_{2,l}$ for, say, odd $l$.
This transformation also yields another duality relations 
between order parameters;
for instance, the $(q_x,\pi)$-spin order parameter 
$\mathcal{O}_{\rm s} (q_x, \pi)$
is the dual of the vector-chirality one 
$\mathcal{O}_{\rm vc}(q_x + \pi)$
with a momentum shift by $\pi$.
We call this transformation the duality {\it II}.

Applying the duality transformation {\it II} 
to the extended model $\mathcal{H}_{\rm ext}$ 
causes a sign change on the $Y$-term 
while the other terms are unchanged. 
The duality {\it II}  therefore yields a mapping 
from $( J_{\rm r}, J_{\rm rr}, W, X, Y, Z, h)$ 
to $( J_{\rm r}, J_{\rm rr}, W, X, -Y, Z, h)$.
The mapping of the original coupling parameters is given by
\begin{eqnarray}
\bar{J}_{\rm r}&=& J_{\rm r},~~~~~\bar{J}_{\rm rr}=J_{\rm rr},
\nonumber \\
\bar{J}_{\rm l} &=& \frac{1}{2}(J_{\rm l}+J_{\rm d})
-\frac{1}{8}(J_{\rm ll}-J_{\rm dd}),
%\nonumber \\
~~~~~
\bar{J}_{\rm d} = \frac{1}{2}(J_{\rm l}+J_{\rm d})
+\frac{1}{8}(J_{\rm ll}-J_{\rm dd}),
\nonumber \\
\bar{J}_{\rm ll}&=& -2(J_{\rm l}-J_{\rm d})
+\frac{1}{2}(J_{\rm ll}+J_{\rm dd}),
%\nonumber \\
~~~~~
\bar{J}_{\rm dd} = 2(J_{\rm l}-J_{\rm d}) 
+ \frac{1}{2}(J_{\rm ll}+J_{\rm dd}).
\end{eqnarray}
The extended model $\mathcal{H}_{\rm ext}$ 
in the parameter surface $Y = 0$, i.e., 
$4(J_{\rm l} -J_{\rm d}) = -(J_{\rm ll} -J_{\rm dd})$
is thus self-dual under the duality {\it II}.

The duality mappings {\it I} and {\it II} enable us 
to make many exact statements on the phase diagram and 
the properties of the model.
For example, if one knows properties of the model 
at a certain parameter point, 
one can immediately translate them to its three dual points.
The mappings also yield a strong constraint on the phase diagram 
that the phase boundaries are symmetric with respect to 
the self-dual surfaces.
Further, if there occurs a direct phase transition between 
two phases which are dual to each other, 
the transition line is exactly on the self-dual surfaces.
Note that a self-dual phase, whose order parameter is self-dual 
under duality transformations {\it I} and/or {\it II}, 
can extend over the corresponding self-dual surfaces.

When one considers the ring-exchange model (\ref{eq:Ham}),
it should be noticed that neither the duality mapping {\it I} nor {\it II} 
preserves the form of the ring-exchange coupling.
As a result, a parameter point on the ring-exchange model
is mapped to a point which is not on the model.
Even so, the duality relations give us useful results. 
The ring-exchange model (\ref{eq:Ham}) is self-dual 
at $J = 2 J_4$ [$\theta = \theta_{\rm sdI} = \tan^{-1}(1/2) 
\sim 0.148 \pi$]\cite{HikiharaMH2003} 
and $J = -2J_4$ [$\theta = \theta_{\rm sdII} 
= \tan^{-1}(-1/2) \sim 0.852 \pi$] 
under the duality mappings {\it I} and {\it II}, respectively.
This is indeed consistent with the phase diagram at $h = 0$ 
obtained in Ref.\ \citen{LaeuchliST2003} :
The self-dual point {\it I}, $J = 2 J_4$, is the transition point 
between the staggered-dimer and scalar-chirality phases, 
which are dual to each other,\cite{HikiharaMH2003} 
while a crossover between the singlet phases with 
the dominant staggered vector-chirality correlation 
and the $(0, \pi)$-spin one 
occurs at the self-dual point {\it II}, $J = -2 J_4$.
Since the duality transformations do not change the magnetic field $h$, 
these dualities give two self-dual lines $J = \pm 2J_4$ 
in the magnetic phase diagram in the $\theta$ versus $h$ plane.
The presence of the self-dual line {\it I} and the fact that 
the model exhibits a TL liquid with a dominant 
$(\pi, \pi)$-transverse-spin quasi-LRO for $\theta = 0$ 
lead us to a prediction:
If the region of the $(\pi, \pi)$-spin-dominant TL liquid extends 
up to the self-dual line {\it I}, 
there must be a crossover to its dual state, i.e., 
a TL liquid with the dominant staggered vector-chirality quasi-LRO, 
exactly at the self-dual line.
We will see in the next section that this is indeed the case.
Further, one may also expect that 
there occurs a crossover at the self-dual line {\it II} 
between the vector-chirality-dominant TL liquid for strong $J_4$ 
and its dual state, $(0, \pi)$-spin-dominant TL liquid, 
for strong ferromagnetic $J$.
As will be shown, this crossover sets in for a small field $h$ 
while for a strong field 
a nematic phase, which is self-dual under the duality {\it II}, 
emerges and extends over the self-dual line {\it II}.

\section{Numerical results}\label{sec:numerics}
\begin{figure*}[tb]
\begin{center}
\includegraphics[width=7.5cm]{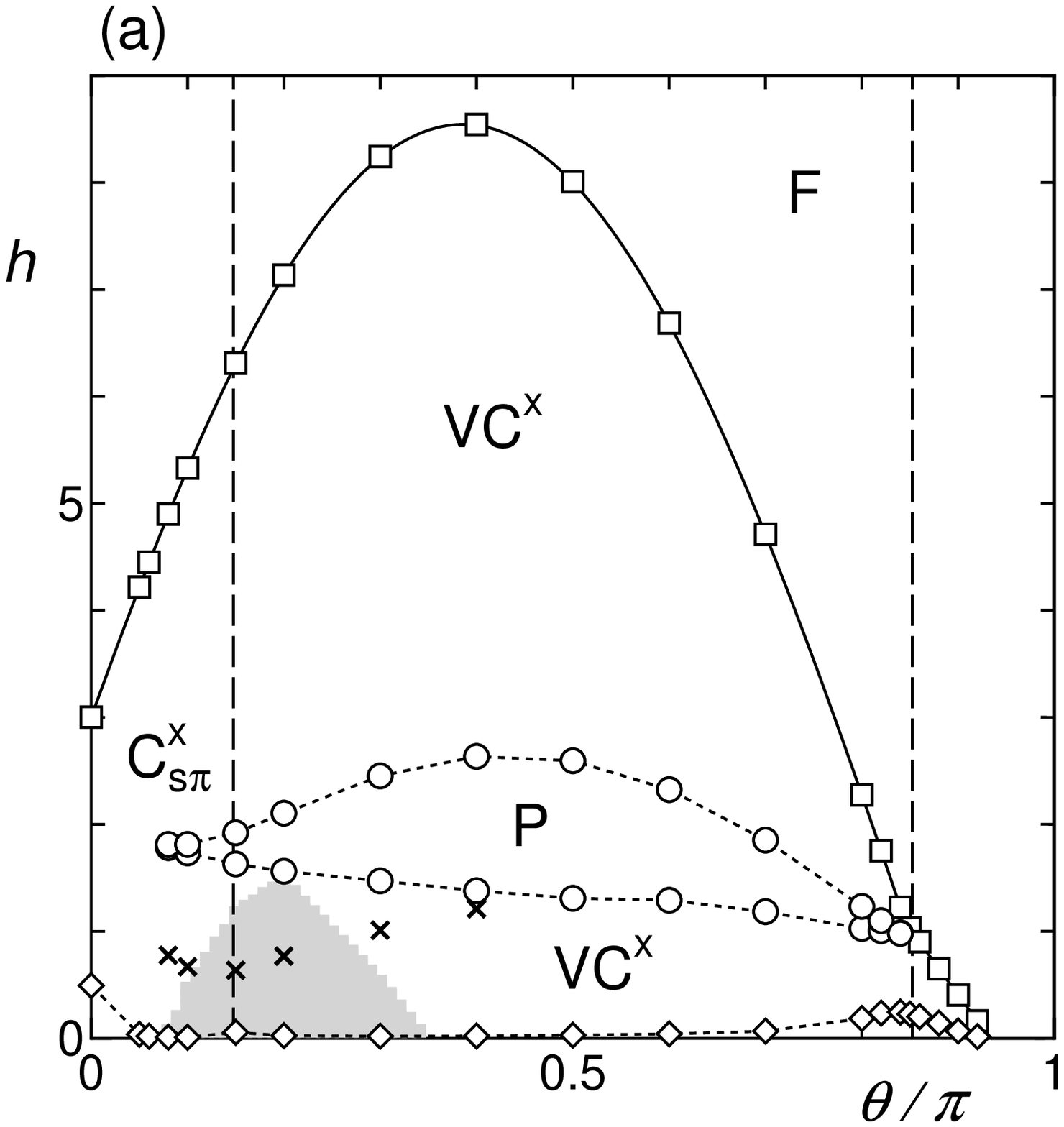}
\includegraphics[width=7.5cm]{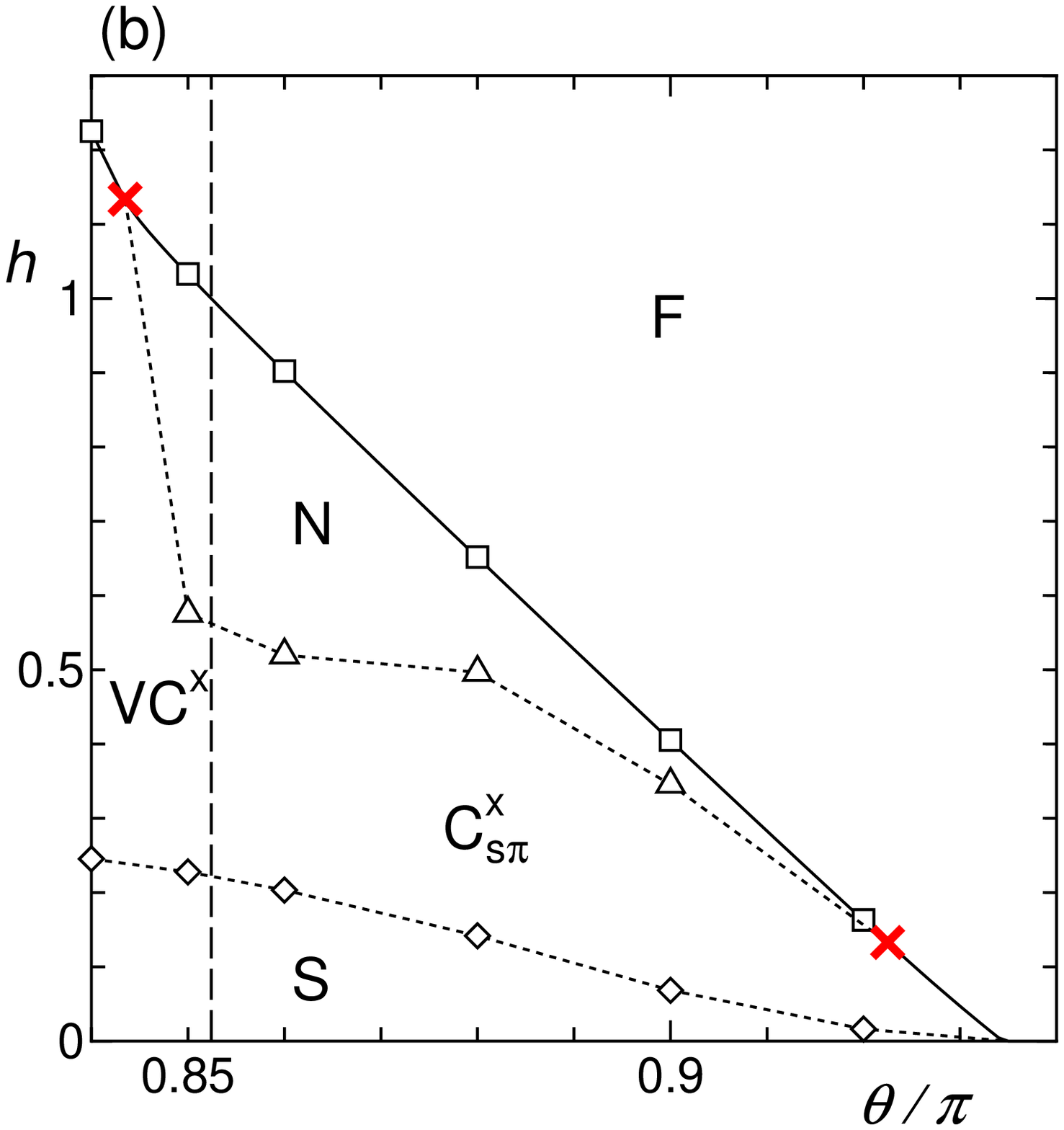}
\end{center}
\caption{
(a) Magnetic phase diagram of the ring-exchange model (\ref{eq:Ham}) 
for $0 \le \theta \le \pi$,
where $J_{\rm rung} = J_{\rm leg} = J = \cos \theta$ and $J_4 = \sin \theta$.
Symbols ^^ ^^ P", ^^ ^^ C$^x_{\rm s \pi}$", ^^ ^^ VC$^x$", ^^ ^^ F" denote 
the regions of the magnetization plateau, transverse-spin-dominant TL liquid, 
vector-chirality-dominant TL liquid, and fully-polarized state, respectively.
Circles represent the critical fields of the plateau boundaries, 
$h_{\rm p1}$ and $h_{\rm p2}$, 
while the squares, diamonds, crosses respectively show 
the saturation field $h_{\rm s}$, 
boundaries $h_{\rm c}$ of the phases at zero magnetization, 
and the field $h_{\rm cusp}$ at the cusp singularity.
Solid curve show the saturation field obtained from 
the exact-diagonalization calculation (see text in Sec.\ \ref{subsec:nematic}).
Vertical dashed lines are the self-dual lines {\it I} and {\it II}.
The shaded area around $\theta \sim 0.2 \pi$ and small $h$ 
indicates a region where the dominant correlation is not identified.
Dotted curves are a guide for eye.
(b) Enlarged phase diagram for $0.84 \pi < \theta < 0.94\pi$.
Symbols ^^ ^^ N" and ^^ ^^ S" denote 
the nematic and spin-singlet phases, respectively,
while triangles and diamonds show the boundaries of the phases.
The bold crosses represent the boundaries 
of the region of two-magnon bound pairs, 
$\theta_{\rm mp1}$ and $\theta_{\rm mp2}$.
}
\label{fig:phasediagram}
\end{figure*}

In this section, we present our numerical results 
for the ring-exchange model $\mathcal{H}$ [Eq.\ (\ref{eq:Ham})] 
under a magnetic field.
To determine the ground-state phase diagram, we have calculated 
the magnetization curve and various correlation functions.
Our findings are summarized in Fig.\ \ref{fig:phasediagram}.
The details of the results are discussed in the following.
We first consider the region of $\theta \lesssim 0.84 \pi$, 
where we find the magnetization plateau 
and vector-chirality-dominant TL liquid.
We then show our results on the nematic phase 
found for $\theta \gtrsim 0.84 \pi$.

\subsection{Magnetization plateau}\label{subsec:plateau}
We first discuss the magnetization curve.
To obtain it, we have calculated the lowest energy $E_0(M)$ 
of the Hamiltonian $\mathcal{H}_0$ [Eq.\ (\ref{eq:Ham_0})] 
in a subspace characterized by the total magnetization per rung, 
$M = S^z_{\rm tot} / L_{\rm r}$, 
where $S^z_{\rm tot} = \sum_l (s^z_{1,l} + s^z_{2,l})$
and $L_{\rm r}$ is the number of rungs.
Note that $M = 1$ for the saturated state.
We then determine the ground-state magnetization $M(h)$
by searching for the magnetization $M$ which minimizes 
the energy $E_0(M; h) = E_0(M) - h S^z_{\rm tot}$ for a given field $h$.
The calculation was performed by 
the DMRG method\cite{White1992,White1993,White1996}
for the finite systems with $L_{\rm r} = 31, 41$, and $51$.
(The reason to use the systems with odd $L_{\rm r}$ will be discussed later.)
The open boundary condition was imposed for the sake of 
the efficiency of the DMRG method.

\begin{figure*}[tb]
\begin{center}
\includegraphics[width=7cm]{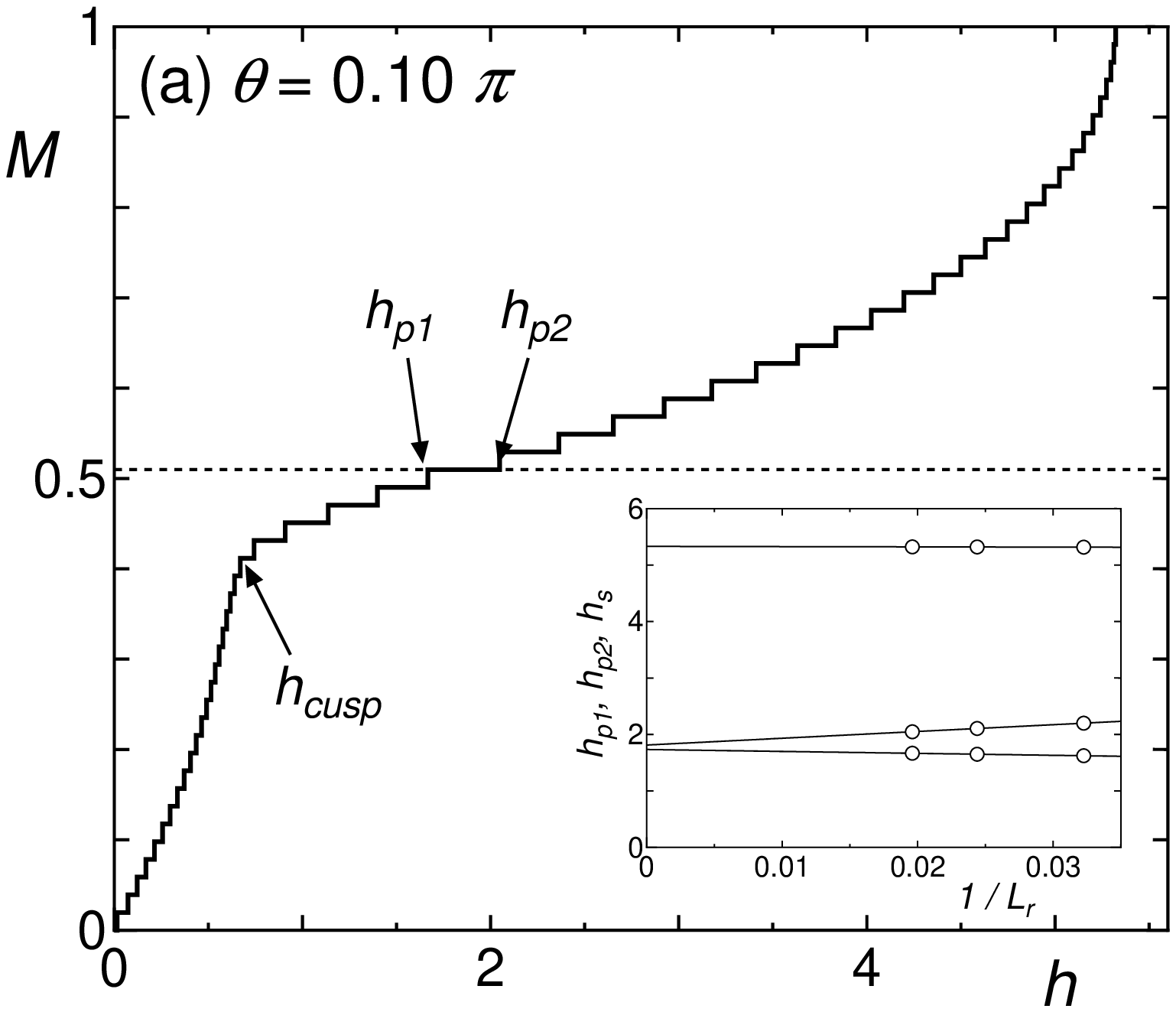}
\includegraphics[width=7cm]{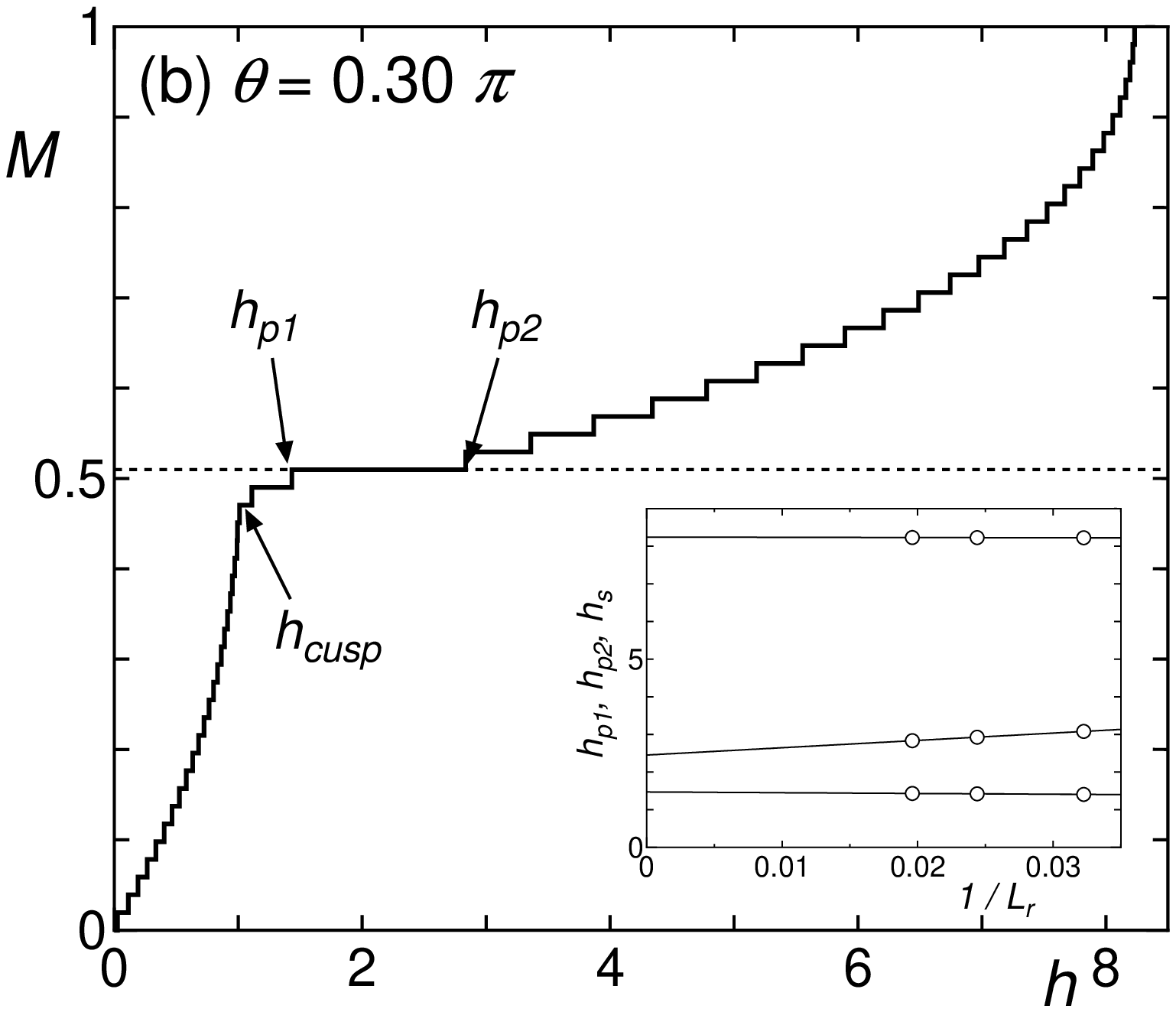}
\includegraphics[width=7cm]{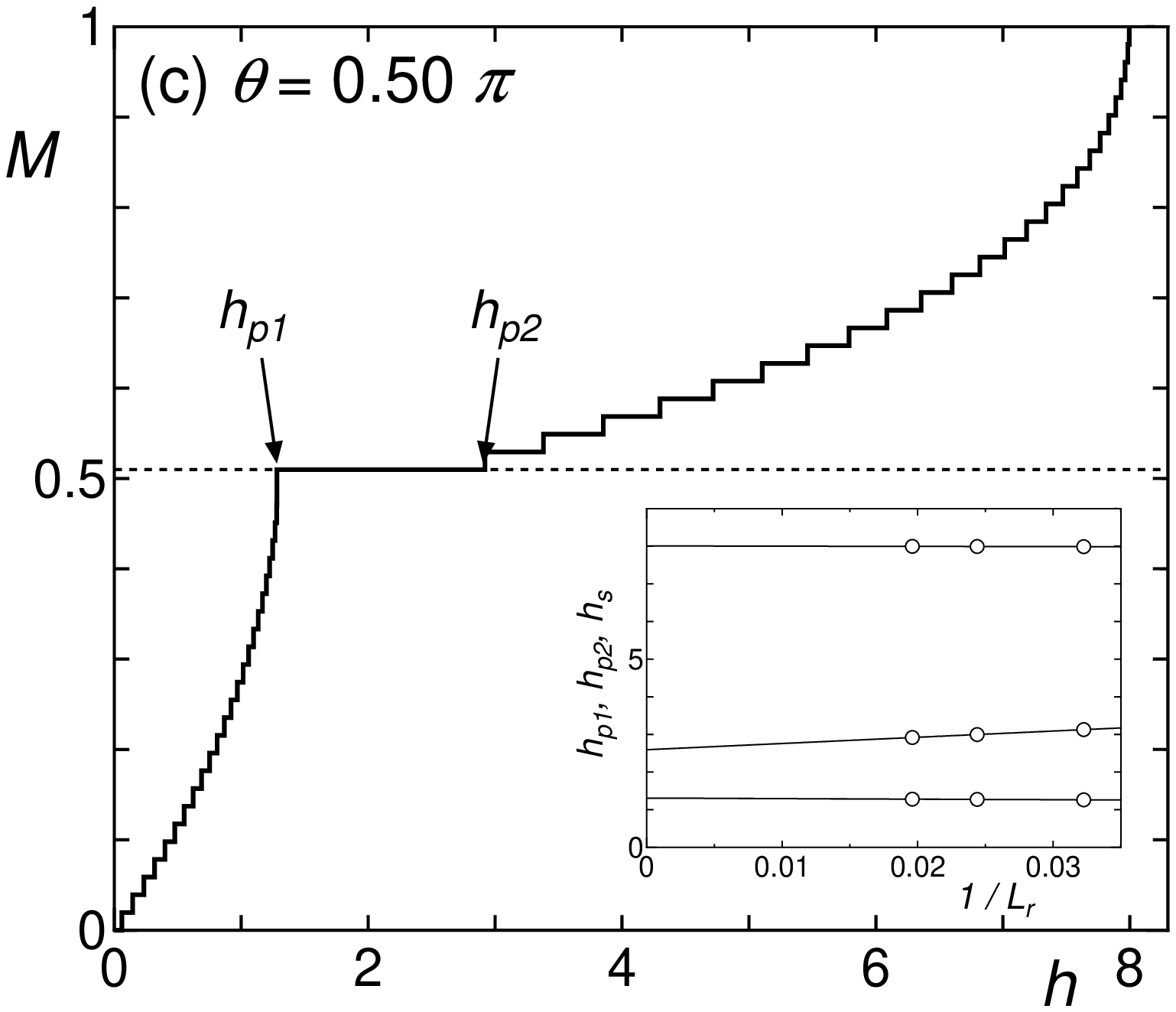}
\includegraphics[width=7cm]{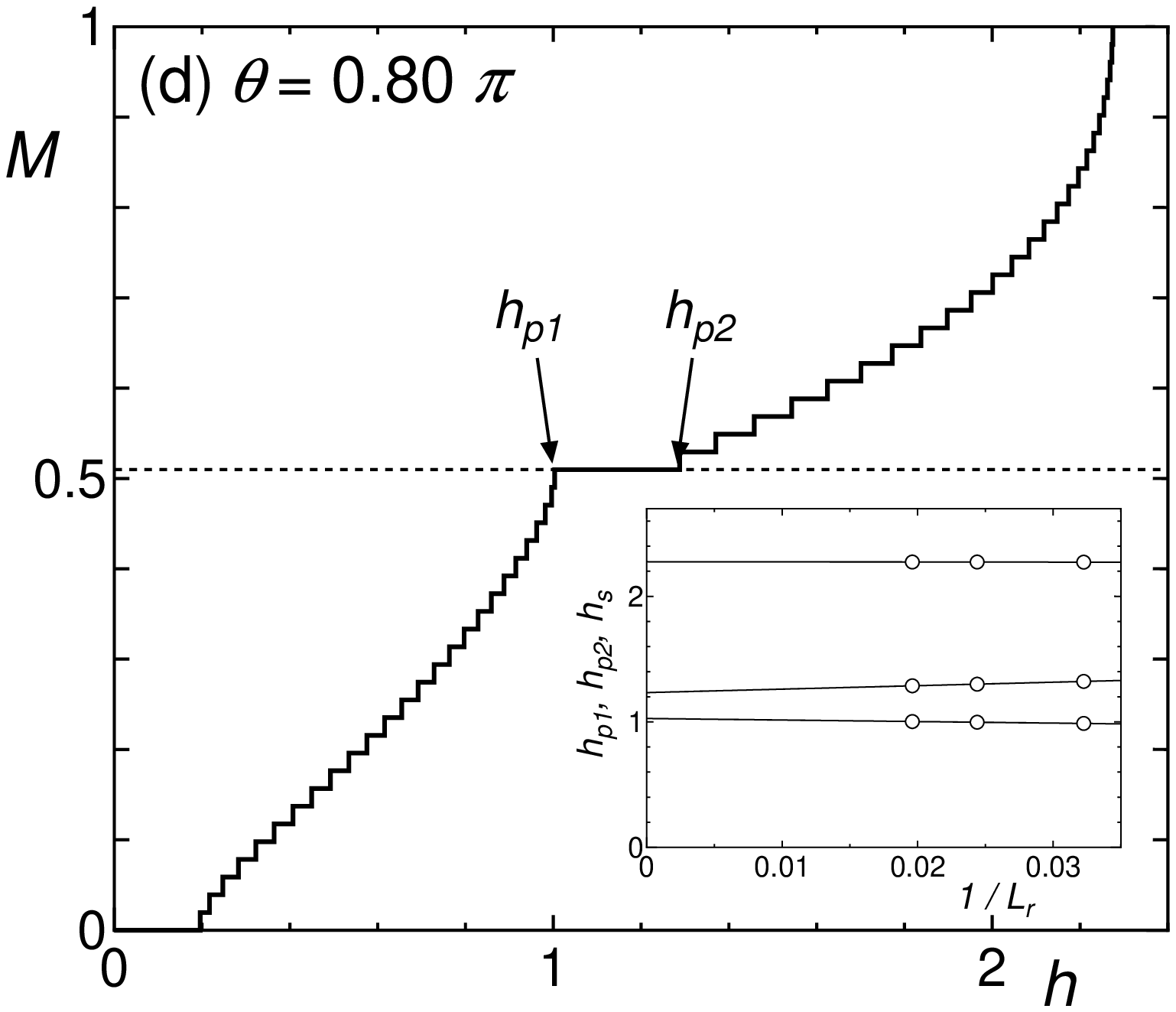}
\end{center}
\caption{
Magnetization curve for (a) $\theta = 0.10 \pi$, (b) $\theta = 0.30 \pi$,
(c) $\theta = 0.50 \pi$, and (d) $\theta = 0.80 \pi$.
The system size is $L_{\rm r} = 51$ and 
the number of the kept states in the DMRG calculation is up to $250$.
The horizontal dotted line represents 
the magnetization at the plateau $M = (L_{\rm r}+1)/(2L_{\rm r})$.
Insets in each panel show the $1/L_{\rm r}$-dependence 
of the fields at the plateau boundaries, $h_{\rm p1}$ and $h_{\rm p2}$,
and the saturated field $h_{\rm s}$.
}
\label{fig:magcurve}
\end{figure*}

\begin{figure}[tb]
\begin{center}
\includegraphics[width=7cm]{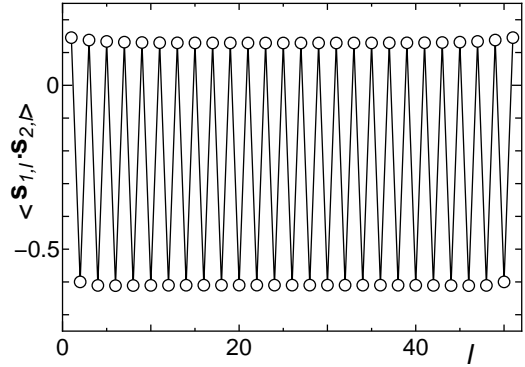}
\end{center}
\caption{
Local spin correlation at each rung, 
$\langle {\bf s}_{1,l} \cdot {\bf s}_{2,l} \rangle$, 
in the magnetization plateau at $\theta = 0.50 \pi$ and $L_{\rm r} = 51$.
}
\label{fig:rungcoupling}
\end{figure}

Figure\ \ref{fig:magcurve} shows the magnetization curves 
for typical values of $\theta$ and $L_{\rm r} = 51$.
It is clearly seen in the figure that 
the magnetization curve exhibits a plateau at the magnetization $M = 1/2$.
[To be precise, for the open ladders with odd $L_{\rm r}$ 
the plateau appears at the magnetization $M = (L_{\rm r}+1)/(2L_{\rm r})$, 
which converges $M \to 1/2$ as $L_{\rm r} \to \infty$.]
After the system-size extrapolation of 
the magnetic field at the upper and lower bounds of the plateau, 
$h_{\rm p1}$ and $h_{\rm p2}$, 
we find that the plateau emerges for a wide range of $\theta$,
$\theta_{\rm p1} < \theta < \theta_{\rm p2}$, 
where the critical values are estimated to be 
$\theta_{\rm p1} = (0.07 \pm 0.01) \pi$ and 
$\theta_{\rm p2} = (0.83 \pm 0.01) \pi$.
The estimate of the lower critical value $\theta_{\rm p1}$ 
is consistent with the previous result $\theta_{\rm p1} = 0.069$
obtained by the exact diagonalization.\cite{NakasuTHOS2001}
The extrapolated values of $h_{\rm p1}$ and $h_{\rm p2}$ 
as well as the saturation field $h_{\rm s}$ 
and the energy gap at zero magnetization $h_{\rm c}$ 
are plotted in the phase diagram Fig.\ \ref{fig:phasediagram}.

It has been predicted by the perturbation theory 
around the limit of the strong rung-coupling\cite{NakasuTHOS2001} that 
the ground state in the plateau has the LRO of the staggered pattern of 
rung-singlet and triplet states.
To confirm this, we have computed 
the ground-state expectation value of 
the correlation between the two spins in each rung, 
$\langle {\bf s}_{1,l} \cdot {\bf s}_{2,l} \rangle$.
The results for $\theta = 0.5 \pi$ 
are shown in Fig.\ \ref{fig:rungcoupling}.
The data clearly demonstrate the appearance of the LRO in the plateau state:
The rungs at the open edges are almost in the spin-triplet state 
and the staggered pattern of triplet and singlet rungs 
penetrates into the bulk without a decay.
Therefore, the translational symmetry is broken spontaneously in the plateau 
and the system has the two-fold degenerate ground states, 
being consistent with the necessary condition 
for the presence of the plateau.\cite{OshikawaYA1997}
We note that one of the two ground states is 
selected in the DMRG calculation due to the open boundaries.
(This is the reason to select odd $L_{\rm r}$ in the analysis 
of magnetization plateau; for even $L_{\rm r}$, there appears a kink 
between two different patterns from the open boundaries 
at the center of the ladder, resulting in an artificial step 
in the plateau of the magnetization curve.)
The appearance of the LRO is observed 
for the whole parameter regime of the magnetization plateau.

Another feature to be noted for the magnetization curve 
is a cusp singularity appearing at a magnetization $M < 1/2$. 
[See Fig.\ \ref{fig:magcurve} (a) and (b).]
The cusp is found for $\theta_{\rm p1} < \theta < 0.5\pi$, 
which corresponds to the region where the plateau exists and 
the bilinear exchanges are antiferromagnetic.
We note that such a cusp singularity, which suggests a change of 
the structure of the excitation spectrum, is often observed 
in frustrated spin systems.\cite{OkunishiHA1999}
The strength of the field at the cusp, $h_{\rm cusp}$, is also 
plotted in the phase diagram, Fig.\ \ref{fig:phasediagram}.

\subsection{Vector chirality}\label{subsec:vectorchirality}
\begin{figure*}[tb]
\begin{center}
\includegraphics[width=7cm]{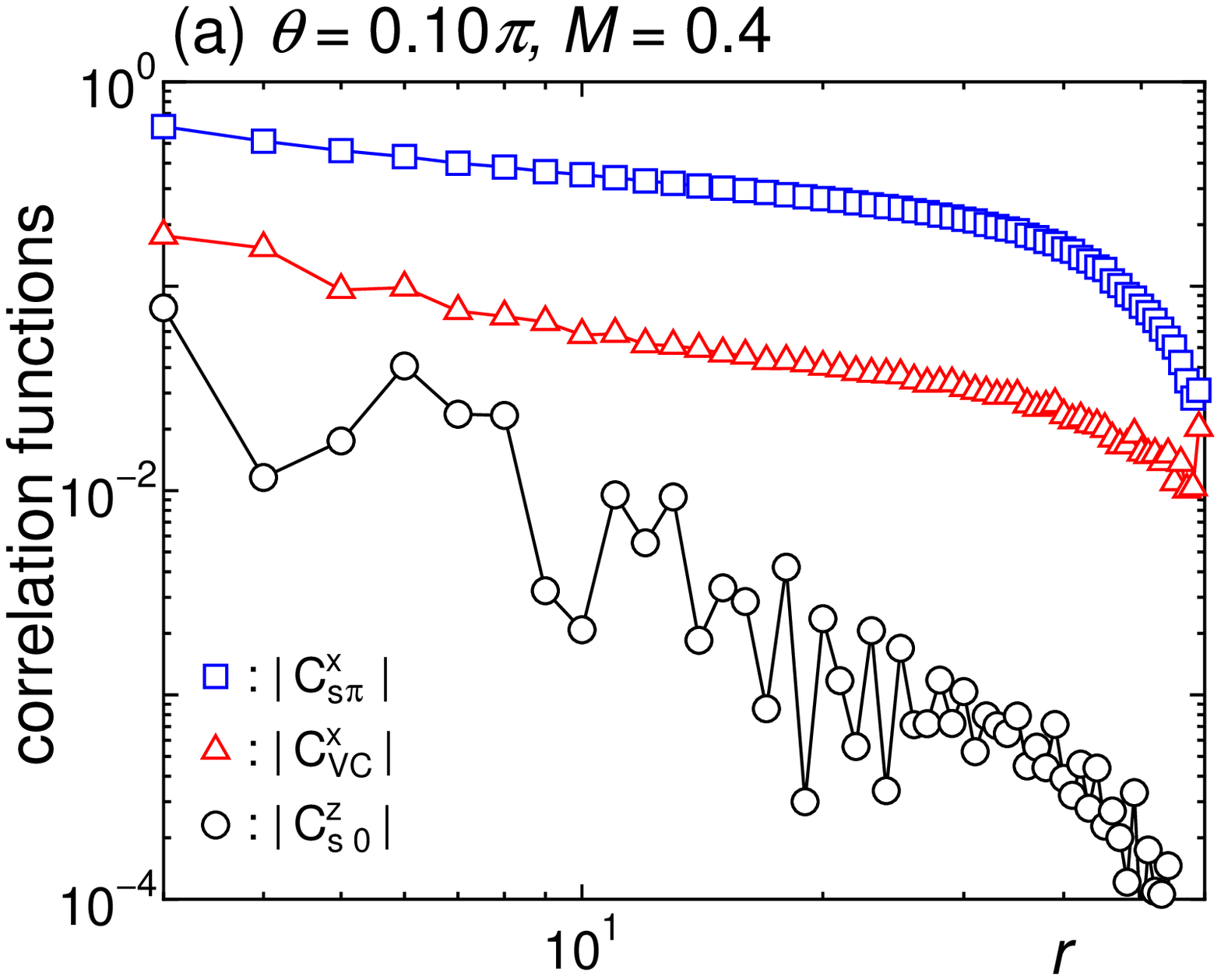}
\includegraphics[width=7cm]{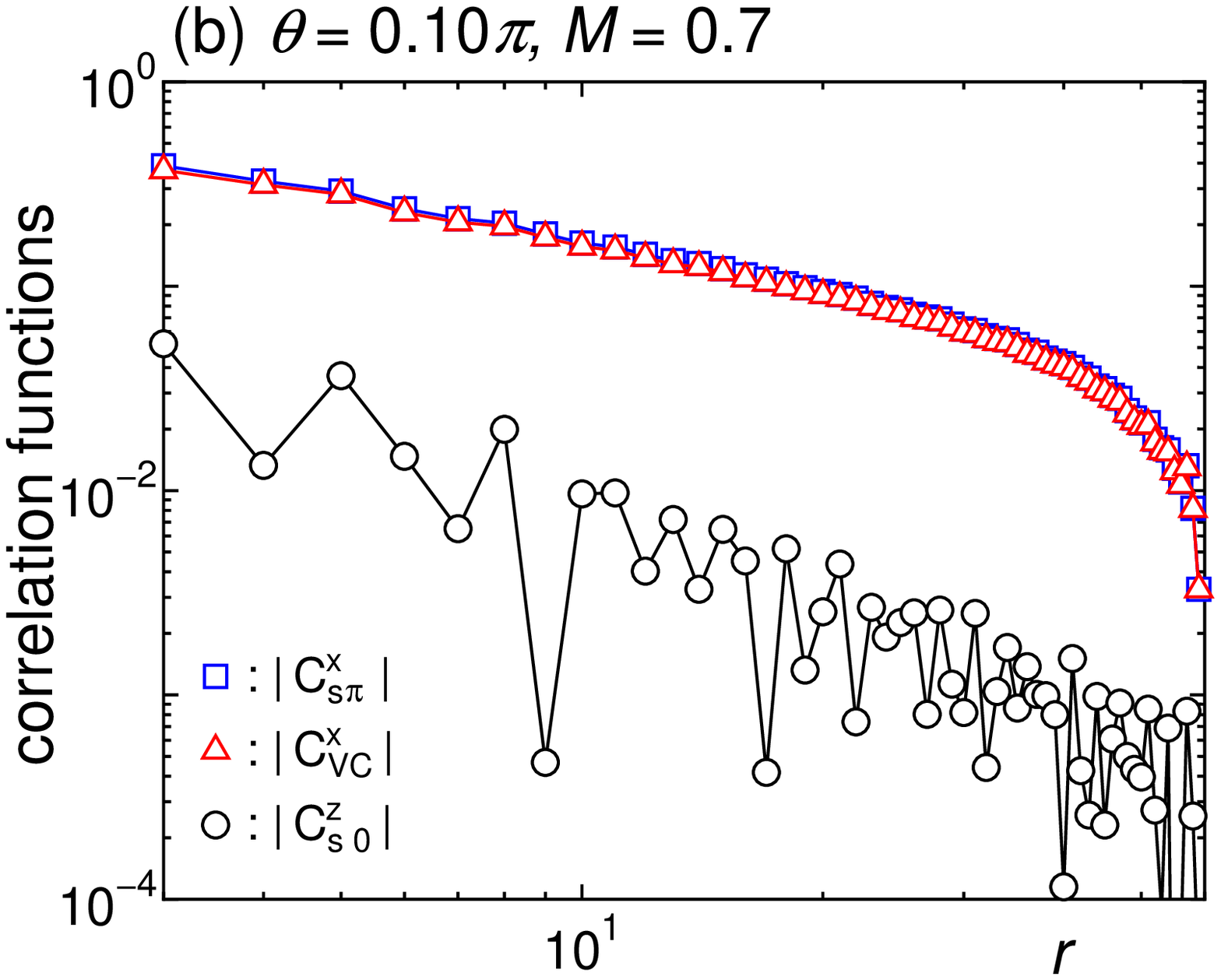}
\includegraphics[width=7cm]{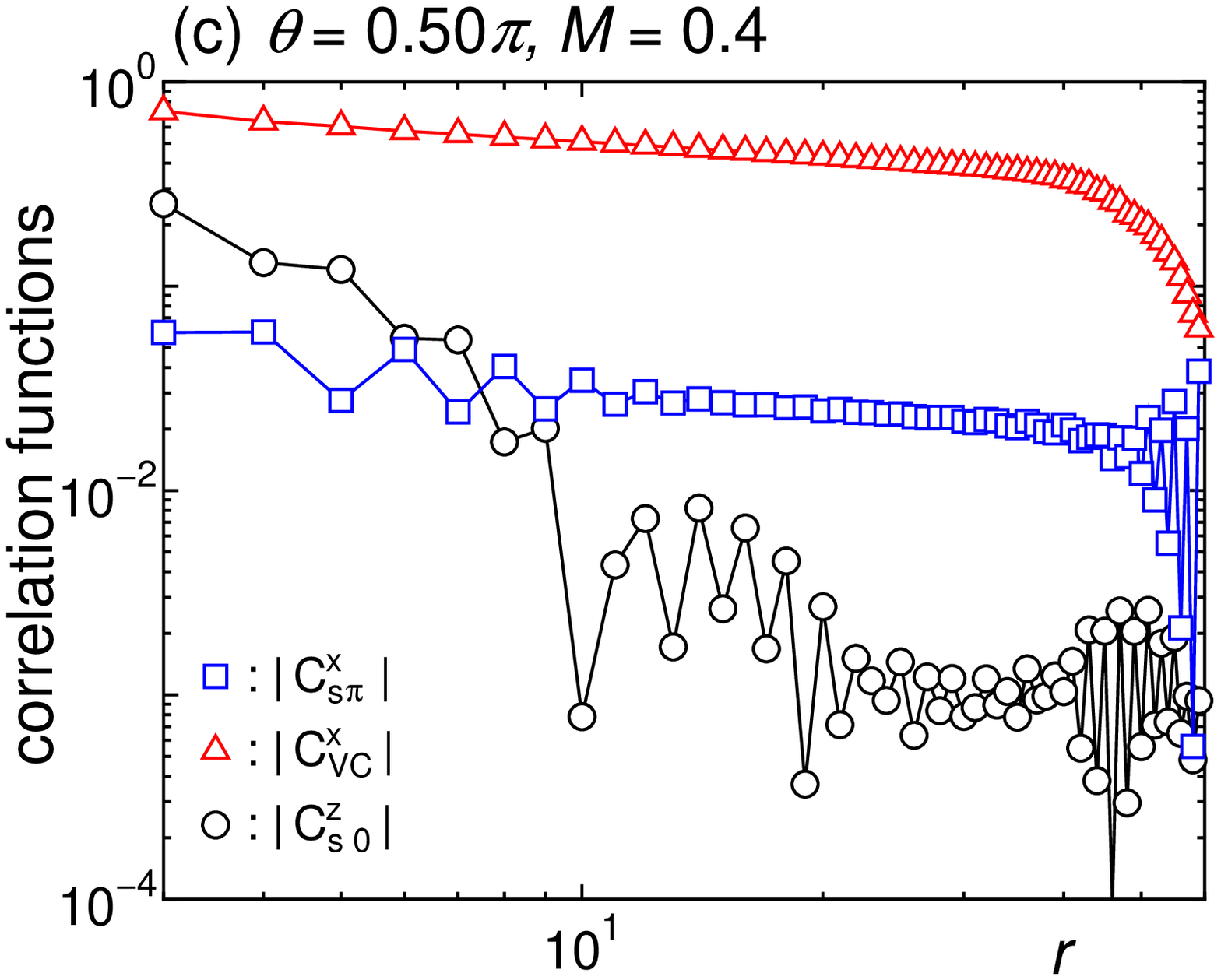}
\includegraphics[width=7cm]{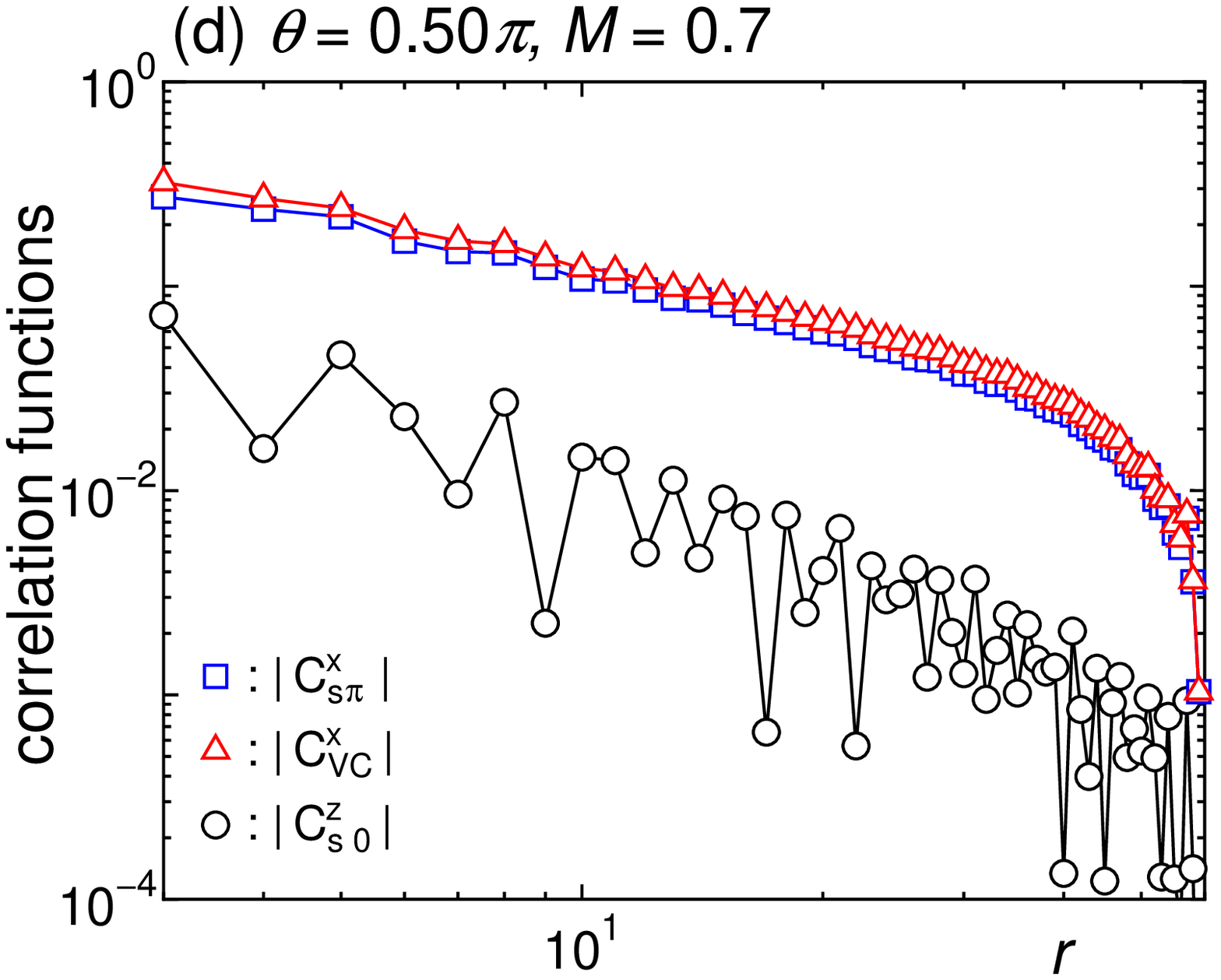}
\end{center}
\caption{
Absolute values of correlation functions 
for (a) $(\theta, M)=(0.1\pi, 0.4)$, (b) $(0.1\pi, 0.7)$,
(c) $(0.5\pi, 0.4)$, and (d) $(0.5\pi, 0.7)$.
Squares, triangles, and circles represent 
the correlation functions $C^x_{\rm s \pi}(r)$, $C^x_{\rm vc}(r)$, 
and $C^z_{\rm s0}(r)$, respectively.
The system size is $L_{\rm r} = 60$ and 
the number of the kept states in the DMRG calculation is up to $300$.
Truncation errors are smaller than the size of the symbols.
}
\label{fig:corVC}
\end{figure*}
\begin{figure}[tb]
\begin{center}
\includegraphics[width=7.5cm]{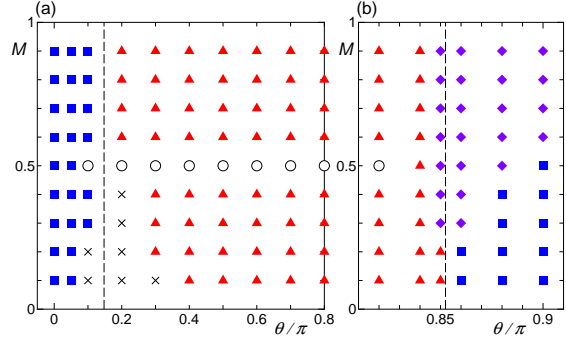}
\end{center}
\caption{
Dominant correlation functions for (a) $0 \le \theta \le 0.8 \pi$ 
and (b) $0.8 \pi < \theta \le 0.9 \pi$.
Squares, triangles, and diamonds represent the critical states 
where the transverse N\'{e}el-type-spin, transverse vector-chirality, 
and magnon-pairing correlation functions are dominant, respectively.
Open circles show the plateau state.
Vertical dashed lines are the self-dual lines {\it I} and {\it II}.
Crosses denote the point where boundary effects are too strong 
to determine the dominant order.
}
\label{fig:dominant}
\end{figure}

Besides the gapped phases of the zero magnetization 
and the magnetization plateau, 
there is a wide region of critical states in the phase diagram.
To characterize the critical states and identify the dominant order, 
we have calculated various two-point correlation functions 
in the ground state of the finite open ladders 
with up to $L_{\rm r} = 60$ rungs using the DMRG method.
The correlation functions considered are 
the ones of total ($q_y=0$) rung spin $s^\alpha_{1,l} + s^\alpha_{2,l}$\ 
($\alpha = x, z$),
N\'{e}el-type ($q_y = \pi$) rung spin $s^\alpha_{1,l} - s^\alpha_{2,l}$, 
vector chirality $({\bf s}_{1,l} \times {\bf s}_{2,l})^\alpha$, 
scalar chirality 
${\bf s}_{1,l} \cdot ({\bf s}_{2,l} \times {\bf s}_{1,l+1})$, 
staggered dimers 
${\bf s}_{1,l} \cdot {\bf s}_{1,l+1}
-{\bf s}_{2,l} \cdot {\bf s}_{2,l+1}$, 
and two-magnon pairing operators 
$s^+_{1,l}s^+_{2,l}$ and $s^+_{1,l}s^+_{1,l+1}$.
To lessen the open boundary effects, 
we calculated the correlation functions on six different pairs 
of two rungs $(l, l')$ for each distance $r = |l-l'|$, 
selecting the center position $l_0 = (l+l')/2$ to be 
as close as possible to the center of the ladder .
We then took the average as an estimate of the correlation function 
for each distance $r$.

Among the correlation functions considered, 
the transverse N\'{e}el-type-rung-spin and 
vector-chirality correlation functions,
\begin{eqnarray}
C^x_{\rm s \pi} (r) 
&=& \frac{1}{S^2} 
\langle (s^x_{1,l} - s^x_{2,l}) (s^x_{1,l'} - s^x_{2,l'}) \rangle,
\label{eq:Cspix} \\
C^x_{\rm vc} (r) 
&=& \frac{1}{S^4} 
\langle ({\bf s}_{1,l } \times {\bf s}_{2,l })^x
        ({\bf s}_{1,l'} \times {\bf s}_{2,l'})^x \rangle,
\label{eq:Cvcx}
\end{eqnarray}
turn out to be dominant in the critical phase 
for $\theta \lesssim 0.84 \pi$.
In Fig. \ \ref{fig:corVC}, we show the data of these correlation functions 
with the longitudinal total-rung-spin one, 
\begin{equation}
C^z_{\rm s 0} (r)  = \frac{1}{S^2} \left[
\langle (s^z_{1,l} + s^z_{2,l}) (s^z_{1,l'} + s^z_{2,l'}) \rangle 
- \langle s^z_{1,l } + s^z_{2,l } \rangle 
  \langle s^z_{1,l'} + s^z_{2,l'} \rangle 
\right].
\label{eq:Cs0z}
\end{equation}
The N\'{e}el-type rung-spin correlation $C^x_{\rm s \pi} (r) $ 
and vector-chirality one $C^x_{\rm vc} (r)$ decay algebraically, 
oscillating with the wavenumber $q_x = \pi$.
Since the decay exponents of these correlations seem to be the same, 
we determine which is the dominant one by comparing their amplitudes.
Figure\ \ref{fig:dominant} summarizes the results 
in the $\theta$ versus $M$ plane.
The N\'{e}el-type rung-spin correlation $C^x_{\rm s \pi} (r)$ 
is dominant for $\theta < \theta_{\rm sdI}$
while the vector-chirality one $C^x_{\rm vc} (r)$ 
is for $\theta > \theta_{\rm sdI}$:
The crossover occurs at the self-dual line {\it I}, 
as discussed in Sec.\ \ref{sec:duality}.
The difference between the amplitudes of 
$C^x_{\rm s \pi} (r)$ and $C^x_{\rm vc} (r)$ 
becomes smaller as the magnetization $M$ increases 
and they converge the same value at $M \to 1$ (see Fig.\ \ref{fig:corVC}).
The vector-chirality-dominant TL liquid thus emerges 
in the wide parameter space of strong ring exchange 
including both the regions of 
the antiferromagnetic and ferromagnetic bilinear exchanges.
We note that there remains a small region 
around $\theta \sim 0.2\pi$ and small $M$ 
where too strong open boundary effects prevent us from 
determining the dominant correlation functions.
For clarifying the nature of the model in the region, 
it is required to perform the calculation for much larger systems, 
which is out of the scope of our numerics 
and left for future studies.

We mention the possible ^^ ^^ $\eta$-inversion" in the system.
For the present model (\ref{eq:Ham}), it was predicted that 
around the plateau there is a parameter regime where 
the longitudinal incommensurate spin correlation function 
becomes stronger than the transverse staggered spin one.\cite{SakaiO2005}
We have found that for $M > 0.5$ the decay exponent of 
the longitudinal total-rung-spin correlation function $C^z_{\rm s0}(r)$
approaches those of $C^x_{\rm s\pi}(r)$ and 
$C^x_{\rm vc}(r)$ as $M$ decreases
and they seem very close to each other at $M = 0.6$.
This seems consistent with the prediction.
However, unfortunately, we do not obtain a clear evidence 
of the phenomenon because of the incommensurate character of 
$C^z_{\rm s0}(r)$ as well as the open boundary effects, 
which make it difficult to estimate the decay exponent 
of $C^z_{\rm s0}$ precisely.
To overcome the difficulties, numerical calculations 
for larger systems and, more preferably, theoretical descriptions 
such as the bosonization analysis for the low-energy states of the system 
would be necessary.

\subsection{Nematic phase}\label{subsec:nematic}
\begin{figure}[tb]
\begin{center}
\includegraphics[width=7.5cm]{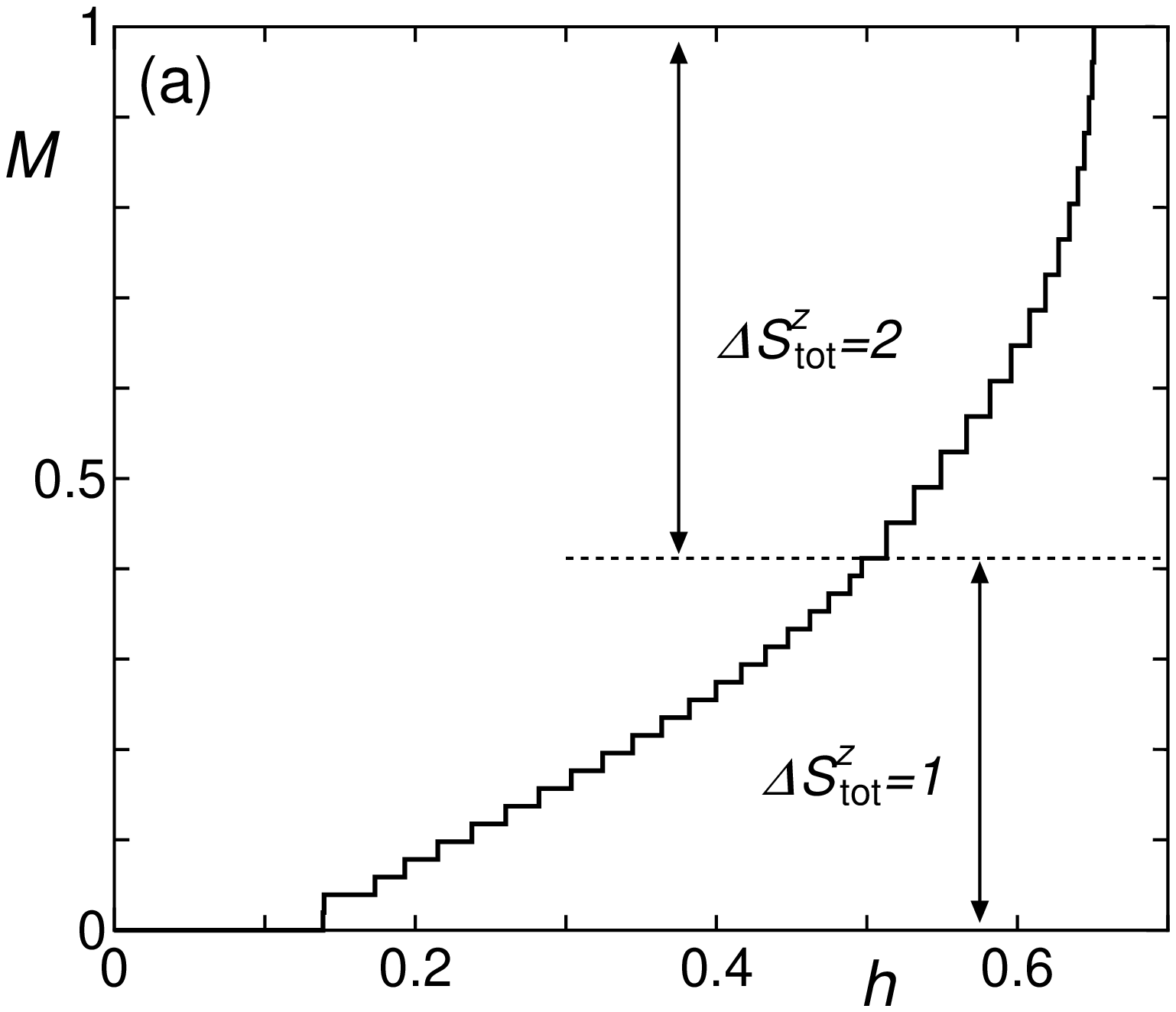}
\includegraphics[width=7.5cm]{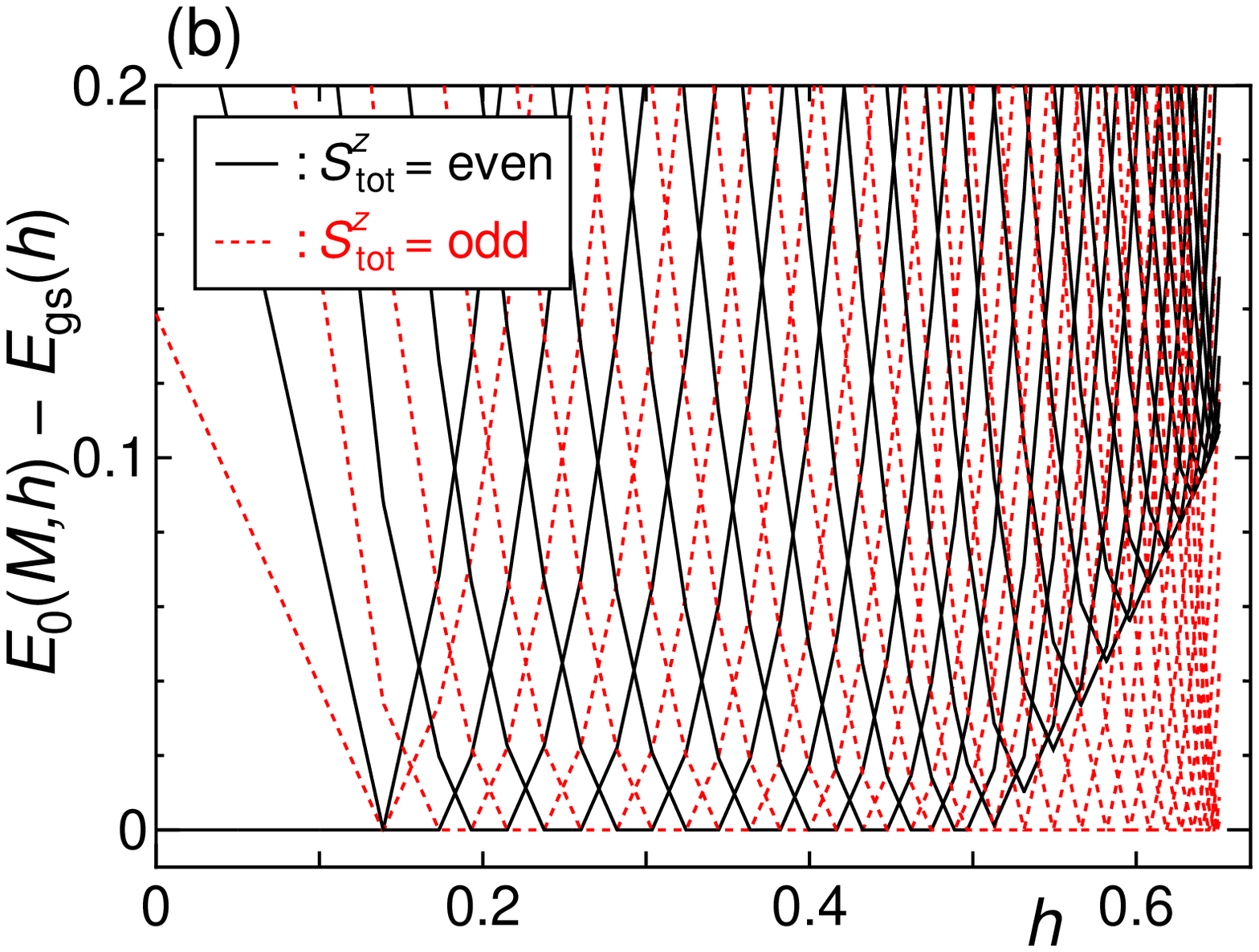}
\end{center}
\caption{
(a) Magnetization curve for $\theta = 0.88 \pi$. 
The system size is $L_{\rm r} = 51$.
(b) Field dependence of excitation energy $E_0(M;h) - E_{\rm gs}(h)$.
Solid and dotted lines represent the energy level 
with even and odd $S^z_{\rm tot}$, respectively.
}
\label{fig:magnonpairing}
\end{figure}
\begin{figure}[tb]
\begin{center}
\includegraphics[width=7.5cm]{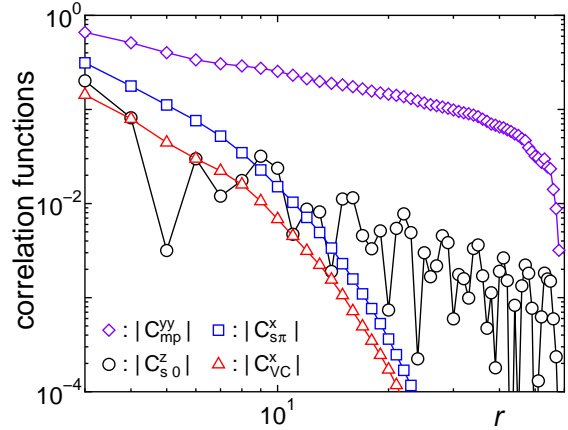}
\end{center}
\caption{
Absolute values of correlation functions 
for $(\theta, M)=(0.88\pi, 0.7)$.
Diamonds, circles, squares, and triangles represent 
the correlation functions 
$C^{yy}_{\rm mp}(r)$, $C^z_{\rm s0}(r)$, 
$C^x_{\rm s \pi}(r)$, and $C^x_{\rm vc}(r)$, respectively.
The system size is $L_{\rm r} = 60$ and 
the number of the kept states in the DMRG calculation is up to $300$.
Truncation errors are smaller 
than the size of the symbols.
}
\label{fig:mpcor}
\end{figure}

Here, we discuss our results for $\theta \gtrsim 0.84 \pi$, 
where we find the nematic phase.
Figure \ref{fig:magnonpairing} (a) shows the magnetization curve 
obtained by the DMRG calculation for $\theta = 0.88 \pi$.
An interesting feature is that 
for a field $h$ larger than a certain critical value 
the magnetization changes by a step $\Delta S^z_{\rm tot} = 2$.
This suggests that magnons in the background of the saturated state 
form bound pairs of two magnons.
To confirm this, we calculate the field dependence of 
the excitation spectrum of the lowest level in each subspace of $M$, 
$E_0(M; h) - E_{\rm gs}(h)$, 
where $E_{\rm gs}(h)$ is the ground-state energy under a field $h$.
The result in Fig.\ \ref{fig:magnonpairing} (b) clearly indicates that 
for large $h$ only the levels with odd $S^z_{\rm tot}$ becomes 
the ground state while the levels with even $S^z_{\rm tot}$ always 
have a finite excitation energy.
We thereby conclude that the magnons indeed form two-magnon bound pairs 
with a finite binding energy.
[Note that, since $L_{\rm r}$ is odd, even (odd) magnons are included 
 in the state with odd (even) $S^z_{\rm tot}$.]

Since the magnetization curve in Fig. \ref{fig:magnonpairing} was 
calculated for the system with the open boundary condition, 
one may think that the step by $\Delta S^z_{\rm tot} = 2$ 
is due to open boundary effects.
To answer the question and determine the boundary of 
the region of the magnon  pairs more accurately, 
we have performed the exact diagonalization calculation 
for the periodic ladder within the subspace of $n$-magnon states, 
where the number of magnons is up to $n = 4$.
The system size treated was up to $L_{\rm r} = 48$, 
which turned out to be large enough.
We calculate the lowest energy in each subspace 
and analyze the magnon instability in the saturated state.
We thereby find that 
for $\theta < \theta_{\rm mp1} = 0.844 \pi$ 
the single-magnon instability is the strongest 
and the saturation field is given by
$h_{\rm s} = 3 J + 8 J_4 = 3 \cos\theta + 8 \sin\theta$, 
which is obtained analytically.
However, when $\theta$ exceeds $\theta_{\rm mp1}$, 
the two-magnon instability dominates and 
determines the saturation field $h_{\rm s}$.
The region of two-magnon bound pairs extends 
for $\theta_{\rm mp1} < \theta < \theta_{\rm mp2} = 0.923 \pi$.
In the small region between $\theta_{\rm mp2}$ and 
the boundary of the ferromagnetic phase $\theta = 0.935\pi$, 
we find that four-magnon instability takes place,
but we must note that the instability of more-magnon bound pairs
may emerge in the vicinity of the ferromagnetic phase.
The saturation field obtained and the boundaries 
of the two-magnon pairing region, $\theta_{\rm mp1}$ and $\theta_{\rm mp2}$,
are plotted in Fig.\ \ref{fig:phasediagram}.

To investigate the nature of the system in the magnon-pairing regime 
in more detail, we have calculated the several correlation functions 
mentioned above using the DMRG method.
From the calculation,
it turns out that the magnon-pairing correlation functions,
\begin{eqnarray}
C^{yy}_{\rm mp} (r) 
&=& \frac{1}{S^4} 
\langle s^+_{1,l}s^+_{2,l} s^-_{1,l'}s^-_{2,l'} \rangle,
\label{eq:Cmpyy} \\
C^{xx}_{\rm mp} (r) 
&=& \frac{1}{S^4} 
\langle s^+_{1,l}s^+_{1,l+1} s^-_{1,l'}s^-_{1,l'+1} \rangle,
\label{eq:Cmpxx} \\
C^{xy}_{\rm mp} (r) 
&=& \frac{1}{S^4} 
\langle s^+_{1,l}s^+_{1,l+1} s^-_{1,l'}s^-_{2,l'} \rangle,
\label{eq:Cmpxy}
\end{eqnarray}
dominate in the region where the two-magnon bound pairs are found.
A typical example of the data of $C^{yy}_{\rm mp} (r)$ 
are shown in Fig.\ \ref{fig:mpcor} with 
the longitudinal total-rung-spin correlation $C^z_{\rm s 0} (r)$, 
the transverse N\'{e}el-type rung-spin one $C^x_{\rm s \pi} (r)$, 
and the transverse vector-chirality one $C^x_{\rm vc} (r)$. 
The magnon-pairing correlation function $C^{yy}_{\rm mp} (r)$ 
is clearly the strongest: $C^z_{\rm s 0} (r)$ decays algebraically 
but faster than $C^{yy}_{\rm mp} (r)$ 
while $C^x_{\rm s \pi} (r)$ and $C^x_{\rm vc} (r)$ decay exponentially.
We note that the other magnon-pairing correlation functions 
$C^{xx}_{\rm mp} (r)$ and $C^{xy}_{\rm mp} (r)$  
also decay algebraically with the same exponent as 
that of $C^{yy}_{\rm mp} (r)$ while the amplitudes are different.
Interestingly, the magnon-pairing correlation functions have 
the sign $C^{yy}_{\rm mp} (r) > 0$, $C^{xx}_{\rm mp} (r) > 0$, 
and $C^{xy}_{\rm mp} (r) < 0$ for arbitrary distance $r$; 
they exhibit the ^^ ^^ $d$-wave-like" symmetry.
All these behaviors are in common with those of the nematic phase 
found for the ring-exchange model 
in the square lattice,\cite{ShannonMS2006} 
in which phase the two-magnon bound pairs undergo a bose condensation 
resulting in the magnon-pairing order, 
though in our one-dimensional system 
the true LRO of the pairing correlation is destroyed 
by a strong quantum fluctuation.
We therefore conclude the ring-exchange ladder (\ref{eq:Ham}) 
for the parameter regime is in (one-dimensional analog of) the nematic phase.
The region of the nematic phase, 
where we find the magnon-bound pairs and 
the dominant magnon-pairing quasi-LRO,
are shown in Figs.\ \ref{fig:phasediagram} (b) and \ref{fig:dominant} (b).
The phase appears at the fields $h$ larger than a critical value $h_{\rm mp}$
and extends over the self-dual line {\it II}. 
This is allowed since the magnon-pairing order parameter is 
self-dual under the duality transformation.
The decay exponents of the magnon-pairing correlation $C^{yy}_{\rm mp}(r)$
and longitudinal total-rung-spin one $C^z_{\rm s 0}(r)$ 
become closer to each other as the magnetization decreases 
and seem almost the same for $\theta = 0.85$ and $M=0.3$.

At a field $h$ lower than the critical value $h_{\rm mp}$, 
we find that the vector-chirality correlation $C^x_{\rm vc} (r)$ 
is dominant for $\theta < \theta_{\rm sdII}$
while the N\'{e}el-type-rung-spin one $C^x_{\rm s \pi} (r)$ 
with wavenumber $q_x = 0$ along the leg direction 
dominates for $\theta > \theta_{\rm sdII}$.
The crossover between the vector-chirality-dominant 
and N\'{e}el-type-spin-dominant TL liquids occurs 
at the self-dual line {\it II}, as expected in Sec.\ \ref{sec:duality}.

\section{Summary and Concluding remarks}\label{sec:summary}
In summary, we have studied the spin-$1/2$ two-leg Heisenberg ladder 
with four-spin ring exchanges under a magnetic field, 
for a wide parameter regime including 
both the antiferromagnetic and ferromagnetic bilinear couplings.
We have introduced a duality transformation, which is 
a simple extension of the spin-chirality duality transformation 
developed previously\cite{HikiharaMH2003,MomoiHNH2003} 
but leads to a nontrivial duality mappings 
on coupling parameters in the Hamiltonian as well as order parameters.
These dualities yield two self-dual lines 
in the parameter space of the ring-exchange ladder (\ref{eq:Ham}).

Further, using the density-matrix renormalization-group and 
exact diagonalization methods, we have determined numerically 
the magnetic phase diagram, Fig.\ \ref{fig:phasediagram},
including the magnetization plateau at $M=1/2$ and 
the regions of TL liquids with different dominant quasi-LRO.
We have found the vector-chirality-dominant TL liquid 
emerging in a wide parameter regime 
of the strong ring-exchange coupling between the self-dual lines, 
while there appear TL liquids with the dominant transverse spin quasi-LRO
outside of the self-dual lines.
Moreover, we have identified the nematic phase, 
in which the magnons form bound pairs and 
the condensation of the bound pairs leads to 
the dominant magnon-pairing correlation functions, 
in a finite regime in the vicinity of the ferromagnetic phase.

The formation of the magnon bound pairs and their condensation 
in the nematic phase are not peculiar to 
the ring-exchange ladder (\ref{eq:Ham}) 
but are found in various $S=1/2$ systems:
The phenomena have been reported for the ring-exchange model 
in the square lattice\cite{ShannonMS2006}  
and triangular lattice,\cite{MomoiSS2006}
and the zigzag ladder with ferromagnetic nearest-neighbor 
and antiferromagnetic next-nearest-neighbor 
couplings.\cite{Chubukov1991,HMeisnerHV2006,VekuaHMHM2007,KeckeMF2007}
These observations suggest that 
the phenomena are a general feature of 
^^ ^^ frustrated ferromagnets" in which the fully-polarized state is 
destabilized by certain perturbations such as ring exchange 
and/or antiferromagnetic couplings.
Systematic studies to develop theoretical descriptions of the state 
and to clarify the relation between the nematic phases 
in $S=1/2$ and $S \ge 1$ systems would be important
for understanding the phenomena.

For the two-leg ladder with extended four-spin exchange 
$\mathcal{H}_{\rm ext}$ [Eq.\ (\ref{eq:Ham_ext})],
it was predicted in Ref.\ \citen{Sato2007} that 
the model under a magnetic field $h > 0$ could realize 
the true LRO of the longitudinal vector-chirality 
$({\bf s}_{1,l} \times {\bf s}_{2,l})^z$, 
while the author also pointed out that 
the form of the ring-exchange coupling, $J_{\rm rr}=J_{\rm ll}=-J_{\rm dd}$, 
was not suitable for the appearance of the LRO. 
Indeed, we have not observed the appearance 
of the true LRO of the vector-chirality in our calculation 
on the ring-exchange ladder (\ref{eq:Ham}).
Searching for the LRO in a ladder model 
with the generalized four-spin exchanges 
of more suitable form would be an interesting issue.

\section*{Acknowledgment}
The authors thank T.\ Momoi and M.\ Sato for fruitful discussions.
The work was supported by a Grant-in-Aid from the Ministry of Education, 
Culture, Sports, Science and Technology (MEXT) of Japan 
(Grant Nos.~18043003).

\end{document}